\documentclass[conference]{IEEEtran}

\usepackage{cite}

\ifCLASSINFOpdf
\else
\fi
\usepackage{graphicx}
\usepackage{epstopdf}
\usepackage{float}
\usepackage{subfloat}
\usepackage{subfig}

\usepackage{tikz}
\usetikzlibrary{shapes,arrows,automata}
\usetikzlibrary{decorations.markings}

\usepackage{amsmath}
\usepackage{mathtools}
\usepackage{amssymb}
\usepackage{amsfonts}

\usepackage{algorithmic}
\usepackage{algorithm}
\usepackage{algorithmic}

\usepackage{xspace}

\makeatletter
\newcommand*{\etc}{%
    \@ifnextchar{.}%
        {etc}%
        {etc.\@\xspace}%
}
\makeatother

\usepackage{amsmath}
\usepackage{mathtools}
\usepackage{amssymb}

\usepackage{array}
\usepackage{multirow}
\usepackage{dcolumn}
\usepackage{booktabs}

\DeclareMathOperator{\ima}{im}

\begin{document}
%
\title{Construction of the generalized \v{C}ech complex}

\author
{
	\IEEEauthorblockN{Ngoc-Khuyen~Le,~Philippe~Martins,~Laurent~Decreusefond~and~Ana\"\i{}s~Vergne}
	\IEEEauthorblockA{Institut Telecom, TELECOM ParisTech, LTCI
	\\Paris, France
	\\Email: ngoc-khuyen.le, martins, decreuse, anais.vergne@telecom-paristech.fr}
}

\maketitle

\begin{abstract}
In this paper, we introduce a centralized algorithm which constructs the generalized \v{C}ech complex. The generalized \v{C}ech complex represents the topology of a wireless network whose cells are different in size. This complex is useful to address a wide variety of problems in wireless networks such as: boundary holes detection, disaster recovery or energy saving. We have shown that our algorithm constructs the minimal generalized \v{C}ech complex, which satisfies the requirements of these applications, in polynomial time.


\end{abstract}


%
\IEEEpeerreviewmaketitle

\section{Introduction}

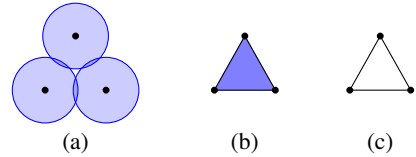
\begin{figure}[width = 0.5\textwidth,H!]
\vspace{-2mm}
\centering
	\begin{tikzpicture}[scale = 0.9, every node/.style={transform shape}]	
	\raggedright
		\begin{scope}[shift={(-0.5,0)},fill opacity = 0.2]
			\fill[blue] (0,0)circle(0.485cm);
			\fill[blue] (0.9,0)circle(0.485cm);
			\fill[blue] (0.45,0.8)circle(0.485cm);
			\filldraw[black,thick,opacity=1] (0,0)circle(1pt);
			\filldraw[black,thick,opacity=1] (0.9,0)circle(1pt);
			\filldraw[black,thick,opacity=1] (0.45,0.8)circle(1pt);
			\draw[blue] (0,0)circle(0.485cm);
			\draw[blue] (0.9,0)circle(0.485cm);
			\draw[blue] (0.45,0.8)circle(0.485cm);
			\draw[black,opacity=1,align=left] (0.45,-0.5)node[below]{(a)};
		\end{scope}
		\begin{scope}[shift={(2,0)},fill opacity = 0.2]
			\filldraw[blue,opacity=0.5] (0,0) -- (0.9,0) -- (0.45,0.8) -- cycle;
			\draw[black,opacity=1] (0,0) -- (0.9,0) -- (0.45,0.8) -- cycle;
			\filldraw[black,thick,opacity=1] (0,0)circle(1pt);
			\filldraw[black,thick,opacity=1] (0.9,0)circle(1pt);
			\filldraw[black,thick,opacity=1] (0.45,0.8)circle(1pt);
			\draw[black,opacity=1,align=left] (0.45,-0.5)node[below]{(b)};
		\end{scope}		
		\begin{scope}[shift={(4,0)},fill opacity = 0.2]
			\draw[black,opacity=1] (0,0) -- (0.9,0) -- (0.45,0.8) -- cycle;
			\filldraw[black,thick,opacity=1] (0,0)circle(1pt);
			\filldraw[black,thick,opacity=1] (0.9,0)circle(1pt);
			\filldraw[black,thick,opacity=1] (0.45,0.8)circle(1pt);
			\draw[black,opacity=1,align=left] (0.45,-0.5)node[below]{(c)};
		\end{scope}
	\end{tikzpicture}	
\hfil\\
\vspace{-1mm}
\caption{(a) Cells, (b) Rips complex, (c) \v{C}ech complex.}
\label{fig:fig_compareRipsCech}
\end{figure}

A wireless network generally contains a group of large number cells. It is interesting to know the topology of the network coverage structure. 
Recent works use simplicial homology to model network coverage.  Indeed, a combinatorial object, named simplicial complex, gives access to the topological information of the network: connectivity and coverage.
Many applications based on simplicial homology have been developed. In \cite{GhristMuhammadCoverageHoleDectection, DeSilvaCoordinateFree, FengConnectivityBasedDist}, some algorithms have been designed in both centralized and decentralized way to locate the coverage holes. In \cite{VergneReduction}, the authors proposed an algorithm to turn off redundant cells without changing the topology of the network. Simplicial homology also helps to recover the wireless network after a disaster \cite{vergne:hal-00800520}. 
These algorithms always need a constructed simplicial complex which represents network coverage structure as their input. 
Concerning simplicial complex, there are two complexes frequently used: the Rips complex and the \v{C}ech complex. The Rips complex represents a group of cells by a simplex if every two of them are neighbors. The Rips complex still describes the neighborhood relation between cells, therefore it sometimes represents inaccurately the topology of the network. The \v{C}ech complex represents a group of cells by a simplex if all of them have a non-empty intersection. If all these cells have the same size, the \v{C}ech complex is called standard. If they are different in size, then this complex is defined as a generalized \v{C}ech complex. The \v{C}ech complex considers the intersection between cells. As a result, it always represents exactly the topology of the network \cite[Theorem 1]{GhristMuhammadCoverageHoleDectection}. 
In Figure \ref{fig:fig_compareRipsCech}, there are three cells with a coverage hole inside them. This hole is represented by an empty triangle in the \v{C}ech representation. However, any two of these cells are neighbor so the Rips complex represents these cells by a filled triangle. It means that there is no coverage hole in the Rips representation. The \v{C}ech complex detects successfully the coverage hole while the Rips complex does not.
In \cite{DantchevConstructionCech}, an algorithm has been proposed to construct the standard \v{C}ech complex. This algorithm, which has been designed to use in computer and graphic science, can only work with a collection of cells which have the same radius. So, this algorithm is not suitable to construct the \v{C}ech complex for the wireless networks whose cells are different in size.\par
\par
In this paper, we introduce an algorithm which constructs the generalized \v{C}ech complex for a collection of cells that are different in size. This algorithm is designed to describe the network coverage structure by a simplicial complex, then one can analyze the coverage structure through it. We also discuss the complexity of our algorithm and then present our simulation results.\par

The rest of this paper is organized as follows. In section \ref{sec:SimplicialHomology}, we introduce the background about simplicial homology and its application in wireless networks. All details of the construction algorithm of the generalized \v{C}ech complex are presented in section \ref{sec:algorithm}. In the next section, the complexity of our algorithm is discussed. Section \ref{sec:experiment} presents and discusses simulation results. Finally, the last section concludes the paper.\par

\section{Simplicial homology and application}
\label{sec:SimplicialHomology}
In this section, we first introduce some notions of simplicial homology. For further details about the simplicial homology, see documents \cite{bookMunkres} and
\cite{bookHatcher}. The application of the simplicial homology in wireless networks is discussed in the latter part of this section.

Given a set of vertices $V$, a $k$-simplex is an
unordered subset $\{v_0, v_1, \ldots, v_k\}$, where $v_i \in V$ and
$v_i \neq v_j$ for all $i \neq j$. The number $k$ is its
dimension. 

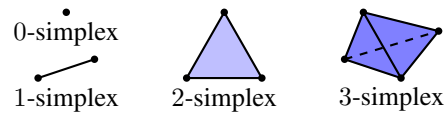
\begin{figure}[H]
\centering
\begin{tikzpicture}[scale = 0.5]
\begin{scope}[shift={(-1.2,0)}]
		\filldraw	[black,thick]	(1,0.75)	circle(2pt)	node[below]{\(0\)-simplex};
		\draw	[black,thick]	(0.25,-1) -- (1.75,-0.50)		(1,-1)node[below]{\(1\)-simplex};
		\filldraw	[black,thick]	(0.25,-1)	circle(2pt);
		\filldraw	[black,thick]	(1.75,-0.50)	circle(2pt);
\end{scope}
\begin{scope}[shift={(0,0)}]
		\filldraw	[blue!25]	(3,-1) -- (5,-1) -- (4,0.75);
		\draw	[black,thick]	(3,-1) -- (5,-1) -- (4,0.75) -- cycle	(4,-1)node[below]{\(2\)-simplex};
		\filldraw	[black,thick]	(3,-1)circle(2pt)	(5,-1)circle(2pt)	(4,0.75)circle(2pt);
\end{scope}
\begin{scope}[shift={(1.2,0)}]
		\filldraw	[blue!50]	(6,-0.5) -- (7.5,-1) -- (8.5,0.25) -- (6.5,0.75);
		\draw	[black,thick]	(6,-0.5) -- (7.5,-1) -- (8.5,0.25) -- (6.5,0.75) -- cycle;	\draw	[black,thick]	(6.5,0.75) -- (7.5,-1)	;	\draw	[black,thick,dashed]	(6,-0.5) -- (8.5,0.25);	\draw	(7.25,-1)node[below]{\(3\)-simplex};
		\filldraw	[black,thick]	(6,-0.5)circle(2pt) (7.5,-1)circle(2pt) (8.5,0.25)circle(2pt) (6.5,0.75)circle(2pt);
\end{scope}
		\end{tikzpicture}
\caption{Examples of simplices.}
\label{fig_simplex}
\end{figure}

Figure \ref{fig_simplex} presents some examples: a $0$-simplex is a point, a $1$-simplex is a
segment of line, a $2$-simplex is a filled triangle, a $3$-simplex is
a filled tetrahedron, \etc.

An oriented simplex is an ordered type of simplex, where swapping
position of two vertices changes its orientation. The change of
orientation is represented by a negative sign as:
\begin{equation*}
[v_0, v_1, \ldots, v_i, v_j, \ldots, v_k] = - [v_0, v_1, \ldots, v_j, v_i, \ldots, v_k]
\end{equation*}

Removing a vertex from a $k$-simplex creates a $(k-1)$-simplex. This
$(k-1)$-simplex is called a face of the $k$-simplex. Thus, each
$k$-simplex has $(k+1)$ faces.

An abstract simplicial complex is a collection of simplices such that: every face of a simplex is also in the simplicial complex.

Let $X$ be a simplicial complex. For each $k \geq 0$, we define a vector
space $C_k(X)$ whose basis is a set of oriented $k$-simplices of
$X$. If $k$ is bigger than the highest dimension of $X$, let $C_k(X) =
0$. We define the boundary operator to be a linear map $\partial : C_k \to
C_{k-1}$ as follows:
\begin{equation*}
\partial[v_0, v_1, \ldots, v_k] = \sum_{i = 0}^{k} (-1)^{i}[v_0, v_1, \ldots, v_{i-1}, v_{i+1}, \ldots, v_k]
\end{equation*}

\begin{figure}[H]
\centering
	\begin{tikzpicture}[scale=0.5, decoration={%
   markings,%
   mark=at position 0.5 with {\arrow[black]{stealth};},%
   }]
	\begin{scope}[shift = {(-1.5,0)}]
		\draw	[postaction=decorate] [black,thick]	(0,0)node[left]{$v_0$} -- (2,0)node[right]{$v_1$};
		\filldraw	[black,thick]	(0,0)	circle(2pt);
		\filldraw	[black,thick]	(2,0)	circle(2pt);
	\end{scope}
	\begin{scope}[shift = {(-1,0)}]
		\draw [black,thick]	(3.5,0)node{$\xrightarrow{\partial}$};	
	\end{scope}
	\begin{scope}[shift = {(0,0)}]
		\filldraw [black,thick]	(5,0)circle(2pt)node[left]{$v_0 +$};
		\filldraw [black,thick]	(7,0)circle(2pt)node[right]{$v_1 -$};
	\end{scope}
	\begin{scope}[shift = {(0.5,0)}]
			\draw [black,thick]	(9,0)node{$\xrightarrow{\partial}$};
	\end{scope}

	\begin{scope}[shift = {(0,0)}]
		\draw (11,0)node{$0$};
	\end{scope}
	\begin{scope}[shift = {(-2,0)}]
		\filldraw	[blue!25]	(0,-3) -- (2,-3) -- (1,-1.25);
		\draw	[black,thick]	(0,-3)node[left]{$v_0$} -- (2,-3)node[right]{$v_1$} -- (1,-1.25)node[above]{$v_2$} -- cycle;
		\filldraw	[black,thick]	(0,-3)circle(2pt) -- (2,-3)circle(2pt) -- (1,-1.25)circle(2pt);
	\end{scope}
	\begin{scope}[shift = {(-1,0)}]
			\draw [black,thick]	(3.5,-2)node{$\xrightarrow{\partial}$};
	\end{scope}
	\begin{scope}[shift = {(0,0)}]
		\draw	[postaction=decorate] [black,thick]	(5,-3)node[left]{$v_0$} -- (7,-3)node[right]{$v_1$};
		\draw	[postaction=decorate] [black,thick] (7,-3)node[right]{$v_1$} -- (6,-1.25)node[above]{$v_2$};
		\draw	[postaction=decorate] [black,thick] (6,-1.25)node[above]{$v_2$} -- (5,-3)node[left]{$v_0$};
		\filldraw	[black,thick]	(5,-3)circle(2pt) -- (7,-3)circle(2pt) --
		(6,-1.25)circle(2pt);
	\end{scope}
	\begin{scope}[shift = {(0.5,0)}]
		\draw [black,thick]	(9,-2)node{$\xrightarrow{\partial}$};
	\end{scope}
	\begin{scope}[shift = {(0,0)}]
		\draw (11,-2)node{$0$};
	\end{scope}
	\begin{scope}[shift = {(-2,0)}]
		\filldraw	[blue!50]	(0,-5.5) -- (1.5,-6) -- (2.5,-4.75) -- (0.5,-4.25);
		\draw	[black,thick]	(0,-5.5) -- (1.5,-6) -- (2.5,-4.75) -- (0.5,-4.25) -- cycle;	\draw	[black,thick]	(0.5,-4.25) -- (1.5,-6)	;	\draw	[black,thick,dashed]	(0,-5			.5) -- (2.5,-4.75);	
		\filldraw	[black,thick]	(0,-5.5)circle(2pt)node[left]{$v_0$} (1.5,-6)circle(2pt)node[below]{$v_1$} (2.5,-4.75)circle(2pt)node[right]{$v_2$} (0.5,-4.25)circle(2pt				)node[above]{$v_3$};
	\end{scope}
	\begin{scope}[shift = {(-1,0)}]
		\draw [black,thick]	(3.5,-5)node{$\xrightarrow{\partial}$};	
	\end{scope} 	
	\begin{scope}[shift = {(0.25,-0.1)}]
		\filldraw	[blue!25, opacity=0.8]	(4.75,-5.75) -- (6.25,-6.25) -- (7.25,-5);
		\draw	[black,thick]	(4.75,-5.75) -- (6.25,-6.25) -- (7.25,-5) -- cycle;
		\filldraw	[black,thick]	(4.75,-5.75)circle(2pt) (6.25,-6.25)circle(2pt) (7.25,-5)circle(2pt);
		
		\filldraw	[blue!25, opacity=0.8]	(4.5,-5.25) -- (7,-4.5) -- (5,-4);
		\draw	[black,thick]	(4.5,-5.25) -- (7,-4.5) -- (5,-4) -- cycle;
		\filldraw	[black,thick]	(4.5,-5.25)circle(2pt) (7,-4.5)circle(2pt) (5,-4)circle(2pt);
		
		\filldraw	[blue!25, opacity=0.8]	(4.5,-5.5) -- (6,-6) -- (5,-4.25);
		\draw	[black,thick]	(4.5,-5.5) -- (6,-6) -- (5,-4.25) -- cycle;
		\filldraw [black,thick] (4.5,-5.5)circle(2pt) (6,-6)circle(2pt) (5,-4.25)circle(2pt);
		
		\filldraw	[blue!25, opacity=0.8]	(6.25,-6) -- (7.25,-4.75) -- (5.25,-4.25);
		\draw	[black,thick]	(6.25,-6) -- (7.25,-4.75) -- (5.25,-4.25) -- cycle;
		\filldraw	[black,thick] (6.25,-6)circle(2pt) (7.25,-4.75)circle(2pt) (5.25,-4.25)circle(2pt);
		\filldraw	[black,thick]	(4.5,-5.5)node[left]{$v_0$};
		\filldraw	[black,thick]	(6.25,-6.25)node[below]{$v_1$};
		\filldraw	[black,thick]	(7.25,-4.75)node[right]{$v_2$};
		\filldraw	[black,thick]	(5,-4)node[above]{$v_3$};
	\end{scope}
	\begin{scope}[shift = {(0.5,0)}]
		\draw [black,thick]	(9,-5)node{$\xrightarrow{\partial}$};
	\end{scope}
	\begin{scope}[shift = {(0,0)}]
		\draw (11,-5)node{$0$};
	\end{scope}
	
	\end{tikzpicture}
	\caption{Boundary operator.}
	\label{fig_boundary}
\end{figure}
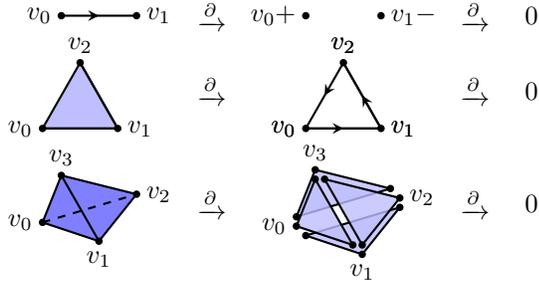

This formula suggests that the boundary of a simplex is the collection of its
faces, as illustrated in Figure \ref{fig_boundary}. For example,
the boundary of a segment is its two endpoints. A filled triangle is
bounded by its three segments. A tetrahedron has its boundary comprised
of its four faces which are four triangles.

\vspace{-1mm}
\vspace{-3mm}

The composition of boundary operators gives a chain of complexes:
\begin{equation*}
\cdots \xrightarrow{\partial{}} C_{k+1} \xrightarrow{\partial{}} C_{k} \xrightarrow{\partial{}} C_{k-1} \cdots \xrightarrow{\partial{}} C_{1} \xrightarrow{\partial{}} C_{0} \xrightarrow{\partial{}} 0 
\end{equation*}

Consider two subspaces of $C_k(X)$: cycle-subspace and
boundary-subspace, denoted as $Z_k(X)$ and $B_k(X)$
respectively. Let $\ker$ be the kernel space and $\ima$ be the image space. By definition, we have:
\begin{equation*}
	\begin{array}{l}
		Z_k(X) = \ker(\partial: C_k \to C_{k-1})\\
		B_k(X) = \ima(\partial: C_{k+1} \to C_{k})\\
	\end{array}
\end{equation*}
$Z_k(X)$ includes cycles which are not boundaries while
$B_k(X)$ includes only boundaries.  A $k$-cycle $u$ is said homologous with a $k$-cycle $v$ if their
difference is a $k$-boundary: $[u] \equiv [v] \Longleftrightarrow u -
v \in B_k(X)$. A simple computation shows that $\partial
\circ \partial = 0$. This result means that a boundary has no
boundary. Thus, the $k$-homology of $X$ is the quotient vector space:
\begin{equation*}
H_k(X) = Z_k(X) \backslash B_k(X)
\end{equation*}

The dimension of $H_k(X)$ is called the $k$-th Betti number:
\begin{equation}
\label{eq:bettiComputation}
\beta{}_k = \dim{H_k} = \dim{Z_k} - \dim{B_k}
\end{equation}

This number has an important meaning for coverage problems. The 
$k$-th Betti number counts the number of $k$-dimensional holes in
a simplicial complex. For example, the $\beta{}_0$ counts the connected
components while $\beta{}_1$ counts the coverage holes,
\etc. \par

\textit{Definition 1 (\v{C}ech complex):} Given $(M,d)$ a metric
space, $\omega$ a finite set of points in $M$ and $\epsilon(\omega)$ a sequence of real
positive numbers, the \v{C}ech complex with parameter $\epsilon(\omega)$ of
$\omega$, denoted $C_{\epsilon(\omega)}(\omega)$ is the abstract simplicial
complex whose $k$-simplices correspond to non-empty intersection of
$(k+1)$ balls of radius $\epsilon(\omega)$ centered at the $(k+1)$ distinct
points of $\omega$.\par

If we choose $\epsilon(\omega)$ to be the cell's coverage range $R$, the \v{C}ech
complex verifies the exact coverage of the system. In the \v{C}ech complex,
each cell is represented by a vertex. A covered space between cells
corresponds to a filled triangle, tetrahedron, \etc. In contrast, a
coverage hole between cells corresponds to an empty (or non-filled)
triangle, rectangle, \etc. \par


\textit{Definition 2 (Index of a vertex):} The index of a vertex $v$ is the biggest integer $k$ such that for every $i \leq k$ each \mbox{$(i-1)$-simplex} of $v$ is a face of at least one $i$-simplex of $v$.\par

The index of a vertex tells us how many times the corresponding cell of this vertex overlaps with its neighbors.
An index of zero indicates that corresponding cell separates from
others (it's isolated). A cell whose index is one still connects to
others by edges. A cell whose index is higher than one connects with
others by triangles, tetrahedron \etc.\par

\begin{figure}[width = 0.5\textwidth,H]
\hspace{8mm}
\subfloat[One filled tetrahedral and one triangle.]{
	\begin{tikzpicture}[scale = 0.75, every node/.style={transform shape}]	
	\raggedright
		\begin{scope}[shift={(0,0)},fill opacity = 0.2]
		\fill[blue!10, very thick] (-1.05,-0.70)rectangle (7.75,1.95);
			\fill[blue]	(0,1.05)circle(0.8cm);		
			\fill[blue]	(-0.1,0.15)circle(0.6cm);		
			\fill[blue]	(0.6,0)circle(0.6cm);		
			\fill[blue]	(1.2,0.2)circle(0.5cm);	
			\fill[blue]	(0.9,0.9)circle(0.8cm);		
			\draw[blue]	(0,1.05)circle(0.8cm);		
			\draw[blue]	(-0.1,0.15)circle(0.6cm);		
			\draw[blue]	(0.6,0)circle(0.6cm);		
			\draw[blue]	(1.2,0.2)circle(0.5cm);	
			\draw[blue]	(0.9,0.9)circle(0.8cm);		
			\filldraw[black,thick,opacity=1]	(0,1.05)circle(1pt)node[left]{0};
			\filldraw[black,thick,opacity=1]	(-0.1,0.15)circle(1pt)node[left]{1};
			\filldraw[black,thick,opacity=1]	(0.6,0)circle(1pt)node[below]{2};
			\filldraw[black,thick,opacity=1]	(1.2,0.2)circle(1pt)node[right]{3};
			\filldraw[black,thick,opacity=1]	(0.9,0.9)circle(1pt)node[above]{4};
		\end{scope}
		\begin{scope}[shift={(3,0)},fill opacity = 0.2]
			\filldraw[blue,opacity=0.5]	(0,1.05) -- (-0.1,0.15) -- (0.6,0) -- (0.9,0.9);
			\filldraw[blue,opacity=0.2] (0.6,0) -- (1.2,0.2) -- (0.9,0.9);
			\draw[black,opacity=1] (0,1.05) -- (-0.1,0.15) -- (0.6,0) -- (0.9,0.9) -- cycle;
			\draw[black,opacity=1] (0,1.05) -- (0.6,0);
			\draw[black,opacity=1,dashed] (-0.1,0.15) -- (0.9,0.9);
			\draw[black,opacity=1] (0.6,0) -- (1.2,0.2) -- (0.9,0.9);
			\filldraw[black,thick,opacity=1]	(0,1.05)circle(1pt)node[left]{0};
			\filldraw[black,thick,opacity=1]	(-0.1,0.15)circle(1pt)node[left]{1};
			\filldraw[black,thick,opacity=1]	(0.6,0)circle(1pt)node[below]{2};
			\filldraw[black,thick,opacity=1]	(1.2,0.2)circle(1pt)node[right]{3};
			\filldraw[black,thick,opacity=1]	(0.9,0.9)circle(1pt)node[above]{4};
		\end{scope}
		\begin{scope}[shift={(4.5,0)}]
			\draw	[black, opacity = 1](0,0.6)node[right]{$
			\begin{array}{l l}
  				\begin{array}{l l}
				    \widehat{i}_0, \widehat{i}_1 = 3 &\\
				    \widehat{i}_2, \widehat{i}_3, \widehat{i}_4 = 2 & \\
				\end{array} 
				&\\
				\begin{array}{l l}
					\beta{}_{0} = 1 &\\
					\beta{}_{1} = 0 &\\
				\end{array}
				&\\	
			\end{array}			
				$};
		\end{scope}
		
	\end{tikzpicture}	
	\label{fig_Cech_a}			
}	
\hfil\\

\hspace{8mm}
\subfloat[Three triangles.]{
	\begin{tikzpicture}[scale = 0.75, every node/.style={transform shape}]
	\raggedright
		\begin{scope}[shift={(0,0)},fill opacity = 0.2]
		\fill[blue!10, very thick] (-1.05,-0.70)rectangle (7.75,1.8);
			\fill[blue]	(0,1.05)circle(0.5cm);		
			\fill[blue]	(-0.1,0.15)circle(0.6cm);		
			\fill[blue]	(0.6,0)circle(0.6cm);		
			\fill[blue]	(1.2,0.2)circle(0.5cm);	
			\fill[blue]	(0.9,0.9)circle(0.8cm);		
			\draw[blue]	(0,1.05)circle(0.5cm);		
			\draw[blue]	(-0.1,0.15)circle(0.6cm);		
			\draw[blue]	(0.6,0)circle(0.6cm);		
			\draw[blue]	(1.2,0.2)circle(0.5cm);	
			\draw[blue]	(0.9,0.9)circle(0.8cm);		
			\filldraw[black,thick,opacity=1]	(0,1.05)circle(1pt)node[left]{0};
			\filldraw[black,thick,opacity=1]	(-0.1,0.15)circle(1pt)node[left]{1};
			\filldraw[black,thick,opacity=1]	(0.6,0)circle(1pt)node[below]{2};
			\filldraw[black,thick,opacity=1]	(1.2,0.2)circle(1pt)node[right]{3};
			\filldraw[black,thick,opacity=1]	(0.9,0.9)circle(1pt)node[above]{4};
		\end{scope}
		\begin{scope}[shift={(3,0)},fill opacity = 0.2]
			\filldraw[blue,opacity=0.2] (0,1.05) -- (-0.1,0.15) -- (0.6,0) -- 					(1.2,0.2) -- (0.9,0.9);
			\draw[black,opacity=1] (0,1.05) -- (-0.1,0.15) -- (0.6,0) -- 						(1.2,0.2) -- (0.9,0.9) -- cycle;
			\draw[black,opacity=1] (0.6,0) -- (0.9,0.9) -- (-0.1,0.15);
			
			\filldraw[black,thick,opacity=1]	(0,1.05)circle(1pt)node[left]{0};
			\filldraw[black,thick,opacity=1]	(-0.1,0.15)circle(1pt)node[left]				{1};
			\filldraw[black,thick,opacity=1]	(0.6,0)circle(1pt)node[below]{2};
			\filldraw[black,thick,opacity=1]	(1.2,0.2)circle(1pt)node[right]					{3};
			\filldraw[black,thick,opacity=1]	(0.9,0.9)circle(1pt)node[above]					{4};
		\end{scope}
		\begin{scope}[shift={(4.5,0)}]
			\draw	[black, opacity = 1](0,0.6)node[right]{$
			\begin{array}{l l}
  				\begin{array}{l l}
				    \widehat{i}_0, \widehat{i}_2, \widehat{i}_1, \widehat{i}_3, \widehat{i}_4 = 2 &\\
				\end{array} &\\
				\begin{array}{l l}
					\beta{}_{0} = 1 &\\
					\beta{}_{1} = 0 &\\
				\end{array}
				&\\	
			\end{array}			
				$};
			
		\end{scope}
	\end{tikzpicture}	
	\label{fig_Cech_b}						
}	
\hfil\\

\hspace{8mm}
\subfloat[One empty hole and one triangle.]{
	\begin{tikzpicture}[scale = 0.75, every node/.style={transform shape}]
	\raggedright
		\begin{scope}[shift={(0,0)},fill opacity = 0.2]
			\fill[blue!10, very thick] (-1.05,-0.7)rectangle (7.75,1.65);
			\fill[blue]	(0,1.05)circle(0.5cm);		
			\fill[blue]	(-0.1,0.15)circle(0.6cm);		
			\fill[blue]	(0.6,0)circle(0.6cm);		
			\fill[blue]	(1.2,0.2)circle(0.5cm);	
			\fill[blue]	(0.9,0.9)circle(0.5cm);		
			\draw[blue]	(0,1.05)circle(0.5cm);		
			\draw[blue]	(-0.1,0.15)circle(0.6cm);		
			\draw[blue]	(0.6,0)circle(0.6cm);		
			\draw[blue]	(1.2,0.2)circle(0.5cm);	
			\draw[blue]	(0.9,0.9)circle(0.5cm);		
			\filldraw[black,thick,opacity=1]	(0,1.05)circle(1pt)node[left]{0};
			\filldraw[black,thick,opacity=1]	(-0.1,0.15)circle(1pt)node[left]{1};
			\filldraw[black,thick,opacity=1]	(0.6,0)circle(1pt)node[below]{2};
			\filldraw[black,thick,opacity=1]	(1.2,0.2)circle(1pt)node[right]{3};
			\filldraw[black,thick,opacity=1]	(0.9,0.9)circle(1pt)node[above]{4};
		\end{scope}
		\begin{scope}[shift={(3,0)},fill opacity = 0.2]
			\filldraw[blue,opacity=0.2] (0.6,0) -- (1.2,0.2) -- (0.9,0.9);
			\draw[black,opacity=1] (0,1.05) -- (-0.1,0.15) -- (0.6,0) -- (1.2,0.2) -- (0.9,0.9) -- cycle;
			\draw[black,opacity=1] (0.6,0) -- (0.9,0.9);
			\filldraw[black,thick,opacity=1]	(0,1.05)circle(1pt)node[left]{0};
			\filldraw[black,thick,opacity=1]	(-0.1,0.15)circle(1pt)node[left]{1};
			\filldraw[black,thick,opacity=1]	(0.6,0)circle(1pt)node[below]{2};
			\filldraw[black,thick,opacity=1]	(1.2,0.2)circle(1pt)node[right]{3};
			\filldraw[black,thick,opacity=1]	(0.9,0.9)circle(1pt)node[above]{4};
		\end{scope}
		\begin{scope}[shift={(4.5,0)}]
			\draw	[black, opacity = 1](0,0.6)node[right]{$
			\begin{array}{l l}
  				\begin{array}{l l}
				    \widehat{i}_0, \widehat{i}_1, \widehat{i}_2, \widehat{i}_4 = 1 &\\
				    \widehat{i}_3 = 2 & \\
				\end{array} 
				&\\
				\begin{array}{l l}
					\beta{}_{0} = 1 &\\
					\beta{}_{1} = 1 &\\
				\end{array}
				&\\	
			\end{array}			
				$};
			
		\end{scope}
	\end{tikzpicture}
	\label{fig_Cech_c}
}	
\hfil\\
\caption{Cells and \v{C}ech representation.}
\label{fig_Cech}
\end{figure}
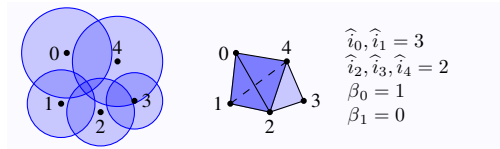
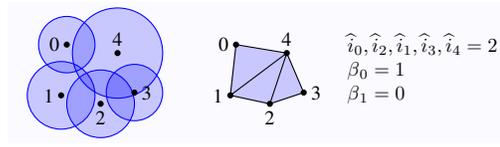
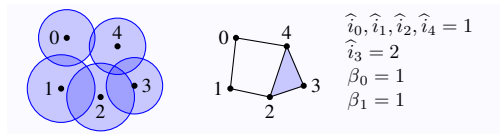

Once the \v{C}ech complex is constructed, we can understand the topology of the network through its homology. The Figure \ref{fig_Cech} shows some examples of cells and their presentation by \v{C}ech complex. In Figure \ref{fig_Cech_a}, four cells 0, 1, 2  and 4 have intersection then they are represented by a tetrahedron. Three cells 2, 3 and 4 also have intersection, so they are represented by a triangle. The Betti number $\beta_0$ is 1 and $\beta_1$ is 0. This means all cells are connected and there is no coverage hole inside them. Cell 0, 1 and their neighbors are always connected by a tetrahedron then cell 0 and 1 have index 3. Cell 2 and 4 are connected with cell 3 by a triangle, so cell 2, 3 and 4 only have index 2. In Figure \ref{fig_Cech_b}, each cell is connected with its neigbors by a triangle, so each one has index 2. All cells are connected and there is no coverage hole so $\beta_0$ is 1 and $\beta_1$ is 0. In Figure \ref{fig_Cech_c}, all cells are connected so $\beta_0$ is 1. There is a coverage hole inside cell 0, 1, 2 and 4. This hole is counted and $\beta_1$ is 1 now. Index of cell 0, 1, 2 and 4 is only 1 which indicates that they lie on the boundary of the hole. Cell 3 is connected with cell 2 and 4 by a triangle so it has index 2.

\section{Construction of the \v{C}ech complex}
\label{sec:algorithm}

To construct the \v{C}ech complex, one needs to verify if any group of cells has a non-empty intersection. Obviously, each 0-simplex represents a cell, the set of 0-simplices, denoted $S_0$, is the list of cells. Each 1-simplex represents a pair of overlapping cells. The construction of 1-simplices boils down to
the search of neighbors for each cell as in the Algorithm \ref{alg:1_simplices}.\par

\begin{algorithm}[H]
\caption{Construction of 1-simplices}
\label{alg:1_simplices}
\begin{algorithmic}
\REQUIRE $S_0$ \COMMENT{collection of cells};
	\STATE $S_1 = \emptyset$; \COMMENT{the set of 1-simplices}
	\STATE $N \gets |S_0|$;
	\FOR{$i = 1 \to N - 1$}
		\FOR{$j = i + 1 \to N$}
			\IF{cell(i) intersects cell(j)}
				\STATE add $s = (i,j)$ to $S_1$;
			\ENDIF
		\ENDFOR
	\ENDFOR
	\RETURN {$S_1$}
\end{algorithmic}
\end{algorithm}

The construction of $k$-simplices where $k \geq 2$ is more complex. The rest of this section is devoted to the details of the construction of $k$-simplices where $k \geq 2$. From the definition of the \v{C}ech complex, each $k$-simplex represents a group of $(k+1)$ cells which have a non-empty intersection. The number of combinations of $(k+1)$ cells in all $N$ cells is huge. We should first find out a candidate group that has an opportunity to be a $k$-simplex. Let us assume that $u = \{v_0, v_1, \ldots, v_k\}$ is a $k$-simplex. Then we can deduce that each pair $(v_i, v_j)$, where $0 \leq i \neq j \leq k$, are neighbors. This suggests that $\hat{u} = \{\hat{v}_0, \hat{v}_1, \ldots, \hat{v}_k\}$, where $\hat{v}_1, \ldots, \hat{v}_k$ are neighbors of $\hat{v}_0$, is a candidate of the cell $\hat{v}_0$ to be a $k$-simplex. If one of $\hat{v}_1, \ldots, \hat{v}_k$ is not neighbor of $\hat{v}_0$, then $\{\hat{v}_0, \hat{v}_1, \ldots, \hat{v}_k\}$ can not be a $k$-simplex. We now need to verify if each candidate is a $k$-simplex. Let us denote $x_{ij}$ an $x_{ji}$ the two intersection points of cell $v_i$ with cell $v_j$. Let $\mathbb{X} = \{x_{ij}, x_{ji} | 0 \leq i < j \leq k\}$ be the set of intersection points for the candidate $\hat{u}$. Let us denote $v_{\ast}$ the smallest cell in candidate cells $\hat{v}_0, \hat{v}_1, \ldots, \hat{v}_k$.
There are only three cases possible. The first case is when the smallest cell $v_{\ast}$ is inside the others. We then conclude that $\hat{u} = \{\hat{v}_0, \hat{v}_1, \ldots, \hat{v}_k\}$ is a $k$-simplex.

\begin{figure}[H]
\centering
	\begin{tikzpicture}[scale = 1, every node/.style={transform shape}]	
	\raggedright
	\fill[blue!02, very thick] (-0.8,-0.80)rectangle (5,2.10);
	\begin{scope}[shift={(0,-1)},fill opacity = 0.2]
		\fill[blue]	(1,2.3)circle(0.3cm);	
		\fill[blue]	(0.5,1.8)circle(1.1cm);	
		\fill[blue]	(1.2,1.5)circle(1.2cm);	
		\fill[blue]	(1.5,2)circle(1cm);	
		
		\draw[blue]	(1,2.3)circle(0.3cm);	
		\draw[blue]	(0.5,1.8)circle(1.1cm);	
		\draw[blue]	(1.2,1.5)circle(1.2cm);	
		\draw[blue]	(1.5,2)circle(1cm);	

		\filldraw[black,thick,opacity=1]	(1,2.3)circle(1pt)node[left]{i};
		\filldraw[black,thick,opacity=1]	(0.5,1.8)circle(1pt)node[left]{j};
		\filldraw[black,thick,opacity=1]	(1.2,1.5)circle(1pt)node[below]{m};
		\filldraw[black,thick,opacity=1]	(1.5,2)circle(1pt)node[right]{n};
		
	\end{scope}
	
	\begin{scope}[shift={(2.75,-1)},fill opacity = 0.2]
		\filldraw[blue,opacity=0.5]	(1,2.3) -- (0.5,1.8) -- (1.2,1.5) -- (1.5,2) ;
		\draw[black,opacity=1] (1,2.3) -- (0.5,1.8) -- (1.2,1.5) -- (1.5,2) -- cycle;
		\draw[black,opacity=1] (1,2.3) -- (1.2,1.5);
		\draw[black,opacity=1,dashed] (0.5,1.8) -- (1.5,2);
		
		\filldraw[black,thick,opacity=1]	(1,2.3)circle(1pt)node[above]{i};
		\filldraw[black,thick,opacity=1]	(0.5,1.8)circle(1pt)node[left]{j};
		\filldraw[black,thick,opacity=1]	(1.2,1.5)circle(1pt)node[below]{m};
		\filldraw[black,thick,opacity=1]	(1.5,2)circle(1pt)node[right]{n};
	\end{scope}		
	\end{tikzpicture}	
\hfil\\
	\caption{Smallest cell is inside the other cells.}
	\label{fig:smallest_inside}
\end{figure}
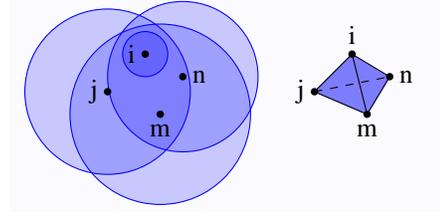

Look at the example from Figure \ref{fig:smallest_inside}, the cell $i$ is the smallest cell and it is inside the others. So the four cells $i,j,m, \text{ and } n$ compose a 3-simplex $(i,j,m,n)$. If the first case is not satisfied, we consider the second case when the smallest cell $v_{\ast}$ is not inside others and there exists an intersection point $x_{ij} \in \mathbb{X}$ that is inside cell $v_{t}$ for all $0 \leq t \leq k$ and $t \neq i,j$.
We then conclude that $\hat{u} = \{\hat{v}_0, \hat{v}_1, \ldots, \hat{v}_k\}$ is a $k$-simplex.

\begin{figure}[H]
\centering
	\begin{tikzpicture}[scale = 1, every node/.style={transform shape}]	
	\raggedright
	\fill[blue!02, very thick] (-0.8,-0.1)rectangle (5,1.9);
	\begin{scope}[shift={(0,-1)},fill opacity = 0.2]
		\fill[blue]	(1,2.3)circle(0.5cm);	
		\fill[blue]	(0.5,1.8)circle(0.65cm);	
		\fill[blue]	(1.2,1.5)circle(0.5cm);	
		\fill[blue]	(1.5,2)circle(0.5cm);	
		\draw[blue]	(1,2.3)circle(0.5cm);	
		\draw[blue]	(0.5,1.8)circle(0.65cm);	
		\draw[blue]	(1.2,1.5)circle(0.5cm);	
		\draw[blue]	(1.5,2)circle(0.5cm);	
		\filldraw[black,thick,opacity=1]	(1,2.3)circle(1pt)node[above]{i};
		\filldraw[black,thick,opacity=1]	(0.5,1.8)circle(1pt)node[left]{j};
		\filldraw[black,thick,opacity=1]	(1.2,1.5)circle(1pt)node[below]{m};
		\filldraw[black,thick,opacity=1]	(1.5,2)circle(1pt)node[right]{n};
		\filldraw[red,thick,opacity=1]	(1.16,1.82)circle(1pt);
	\end{scope}
	
	\begin{scope}[shift={(2.75,-1)},fill opacity = 0.2]
		\filldraw[blue,opacity=0.5]	(1,2.3) -- (0.5,1.8) -- (1.2,1.5) -- (1.5,2) ;
		\draw[black,opacity=1] (1,2.3) -- (0.5,1.8) -- (1.2,1.5) -- (1.5,2) -- cycle;
		\draw[black,opacity=1] (1,2.3) -- (1.2,1.5);
		\draw[black,opacity=1,dashed] (0.5,1.8) -- (1.5,2);
		
		\filldraw[black,thick,opacity=1]	(1,2.3)circle(1pt)node[above]{i};
		\filldraw[black,thick,opacity=1]	(0.5,1.8)circle(1pt)node[left]{j};
		\filldraw[black,thick,opacity=1]	(1.2,1.5)circle(1pt)node[below]{m};
		\filldraw[black,thick,opacity=1]	(1.5,2)circle(1pt)node[right]{n};
	\end{scope}		
	\end{tikzpicture}	
\hfil\\
	\caption{One of the intersection points of cell $i$ and cell $j$ is inside the other cells.}
	\label{fig:one_inside}
\end{figure}
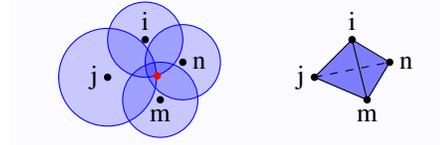

In Figure \ref{fig:one_inside}, an intersection point of cell $i$ and cell $j$ is inside other cells $m$ and $n$. This point is marked as a red point in Figure \ref{fig:one_inside}. So, we conclude that $(i,j,m,n)$ is a 3-simplex. If both first case and second case are not satisfied, we conclude that $\hat{u} = \{\hat{v}_0, \hat{v}_1, \ldots, \hat{v}_k\}$ is not a $k$-simplex. The smallest cell $v_{\ast}$ is not inside others and no intersection point $x_{ij} \in \mathbb{X}$ is inside cell $v_{t}$ for any $0 \leq t \leq k$ and $t \neq i,j$. So there must exist $i^{\ast}$, $j^{\ast}$ and $m^{\ast}$ such that $x_{i^{\ast}}$ and $x_{j^{\ast}}$ are not inside the cell $m^{\ast}$. So $(i^{\ast}, j^{\ast}, m^{\ast})$ is not a 2-simplex then it can not be part of a $k$-simplex with $k \geq 2$. 

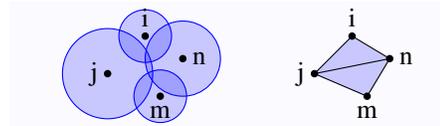
\begin{figure}[H]
\centering
	\begin{tikzpicture}[scale = 1, every node/.style={transform shape}]	
	\raggedright
	\fill[blue!02, very thick] (-0.8,0.05)rectangle (5,1.75);
	\begin{scope}[shift={(0,-1)},fill opacity = 0.2]
		\fill[blue]	(1,2.3)circle(0.35cm);	
		\fill[blue]	(0.5,1.8)circle(0.6cm);	
		\fill[blue]	(1.2,1.5)circle(0.35cm);	
		\fill[blue]	(1.5,2)circle(0.5cm);	
		\draw[blue]	(1,2.3)circle(0.35cm);	
		\draw[blue]	(0.5,1.8)circle(0.6cm);	
		\draw[blue]	(1.2,1.5)circle(0.35cm);	
		\draw[blue]	(1.5,2)circle(0.5cm);	
		\filldraw[black,thick,opacity=1]	(1,2.3)circle(1pt)node[above]{i};
		\filldraw[black,thick,opacity=1]	(0.5,1.8)circle(1pt)node[left]{j};
		\filldraw[black,thick,opacity=1]	(1.2,1.5)circle(1pt)node[below]{m};
		\filldraw[black,thick,opacity=1]	(1.5,2)circle(1pt)node[right]{n};

	\end{scope}
	
	\begin{scope}[shift={(2.75,-1)},fill opacity = 0.2]
		\filldraw[blue,opacity=0.2]	(1,2.3) -- (0.5,1.8) -- (1.2,1.5) -- (1.5,2) ;
		\draw[black,opacity=1] (1,2.3) -- (0.5,1.8) -- (1.2,1.5) -- (1.5,2) -- cycle;
		\draw[black,opacity=1] (0.5,1.8) -- (1.5,2);
		
		\filldraw[black,thick,opacity=1]	(1,2.3)circle(1pt)node[above]{i};
		\filldraw[black,thick,opacity=1]	(0.5,1.8)circle(1pt)node[left]{j};
		\filldraw[black,thick,opacity=1]	(1.2,1.5)circle(1pt)node[below]{m};
		\filldraw[black,thick,opacity=1]	(1.5,2)circle(1pt)node[right]{n};
	\end{scope}		
	\end{tikzpicture}	
\hfil\\
	\caption{No intersection point of any two cells is inside the other cells.}
	\label{fig:non_inside}
\end{figure}

In Figure \ref{fig:non_inside}, we can not find a pair of cells for which one of their intersection points is inside all the other cells. In addition, both intersection points of cell $i$ and cell $j$ are not inside cell $m$. So the three cells $i,j \text{ and } m$ do not compose a 2-simplex. Then, four cells $i,j,m \text{ and } n$ do not compose a 3-simplex. The algorithm to verify if a candidate is a $k$-simplex, where $k \geq 2$, is given in the Algorithm \ref{alg:verification}:

\begin{algorithm}[H]
\caption{Verification a candidate of $k$-simplex}
\label{alg:verification}
\begin{algorithmic}
\REQUIRE $\hat{u} = \{\hat{v}_0, \hat{v}_1, \ldots, \hat{v}_k\}$ a candidate;
	\STATE $v^{\ast} \gets$ the smallest cell of $\{\hat{v}_0, \hat{v}_1, \ldots, \hat{v}_k\}$;
	\IF{cell $v^{\ast}$ is inside cell $\hat{v}_i$ for all $\hat{v}_i \neq v^{\ast}$}
		\STATE verification $= \TRUE$; 
	\ELSE
		\STATE $\mathbb{X} = \emptyset$;
		\FOR{$i = 0 \to k$}
			\FOR{$j = i+1 \to k+1$}
				\STATE $\{x_{ij}, x_{ji}\} = $ intersection points of cells $i$, $j$;
				\STATE add $\{x_{ij}, x_{ji}\}$ into $\mathbb{X}$;
			\ENDFOR
		\ENDFOR
		\IF{there exists $x_{ij}$ inside cell $\hat{v}_t$ for all $t \neq i,j$}
			\STATE verification $= \TRUE$;
		\ELSE
			\STATE verification $= \FALSE$;
		\ENDIF
	\ENDIF
	\RETURN verification;
\end{algorithmic}
\end{algorithm}
\vspace{-1mm}
The construction of the \v{C}ech complex can be summarized as in the Algorithm \ref{alg:construction_ksimplex}:
\vspace{-1mm}
\begin{algorithm}[H]
\caption{Construction of the \v{C}ech complex}
\label{alg:construction_ksimplex}
\begin{algorithmic}
\REQUIRE {$S_0 $ and $S_1$};
\STATE k = 2;
\WHILE{(1)}
	\STATE $S_k = \emptyset$; \% collection of $k$-simplices
	\FOR{each $\hat{v}_0 \in S_0$}
		\STATE $S^{\ast} = $ a set of candidate $\{\hat{v}_0, \hat{v}_1, \ldots, \hat{v}_k\}$ of $\hat{v}_0$;
		\FOR{each $\hat{u} = \{\hat{v}_0, \hat{v}_1, \ldots, \hat{v}_k\} \in S^{\ast}$}
			\STATE verification $= \text{verify}(\hat{u})$; \% call to Algorithm \ref{alg:verification}
			\IF{$\text{verification} = \TRUE$}
				\STATE add $\hat{u}$ into $S_k$;
			\ENDIF
		\ENDFOR
	\ENDFOR
	\IF{$S_k \neq \emptyset$}
		\STATE $k = k+1$;
	\ELSE
		\STATE break;
	\ENDIF
\ENDWHILE
\RETURN sequence of $S_k$ for all $k \geq 2$;
\end{algorithmic}
\end{algorithm}

\section{Complexity}

To construct the \v{C}ech complex, one needs to verify if any group of cells has a non-empty intersection.
The 0-simplices are obviously a collection of vertices.
Computing 1-simplices is to search of neighbors for each cell. Its complexity is $C^2_N$, 
where $N$ is the number of cells. To compute 2-simplices, for each cell we take two 
of its neighbors and verify if this cell and the two neighbors have a non-empty 
intersection. The verification has complexity $O(k^2)$ where $k$ is the dimension of the simplex. This verification can be done in an instant time. Let $n$ be the average number of neighbors of each cell, the complexity
to compute 2-simplices for each cell is $C^2_n$ on average. The complexity of the 2-simplices computation
for all cells is $NC^2_n$. Similarly, to compute the $k$-simplices for each cell, we take $k$ 
of its neighbors and verify if this cell and the neighbors have a non-empty intersection.
The complexity of $k$-simplices computation of one cell is $C^k_n$ and for all cells is 
$NC^k_n$. Consequently, the complexity to construct the \v{C}ech complex is: 
$C^2_N + N\sum_{k=2}^{d_{\text{max}}} C^k_n$, where $d_{\text{max}}$ is the highest
dimension of the \v{C}ech complex. 
Many applications in wireless networks such as locating coverage hole or disaster recovery require only the \v{C}ech complex built up to dimension 2. In this case, the complexity to construct the \v{C}ech complex up to dimension 2 is only $O(N^2 + Nn^2)$. If the \v{C}ech complex is built up to its highest dimension, the sum $\sum_{k=2}^{d_{\text{max}}} C^k_n$ can be upper bounded by $2^n$. The complexity
to construct the \v{C}ech complex up to highest dimension is then as much $O(N^2 + N2^n)$.

\section{Simulation results}
\label{sec:experiment}

\begin{figure}[H]
\centering
\includegraphics[trim = 20mm 6.5mm 20mm 6.5mm, clip, width = 0.35\textwidth]{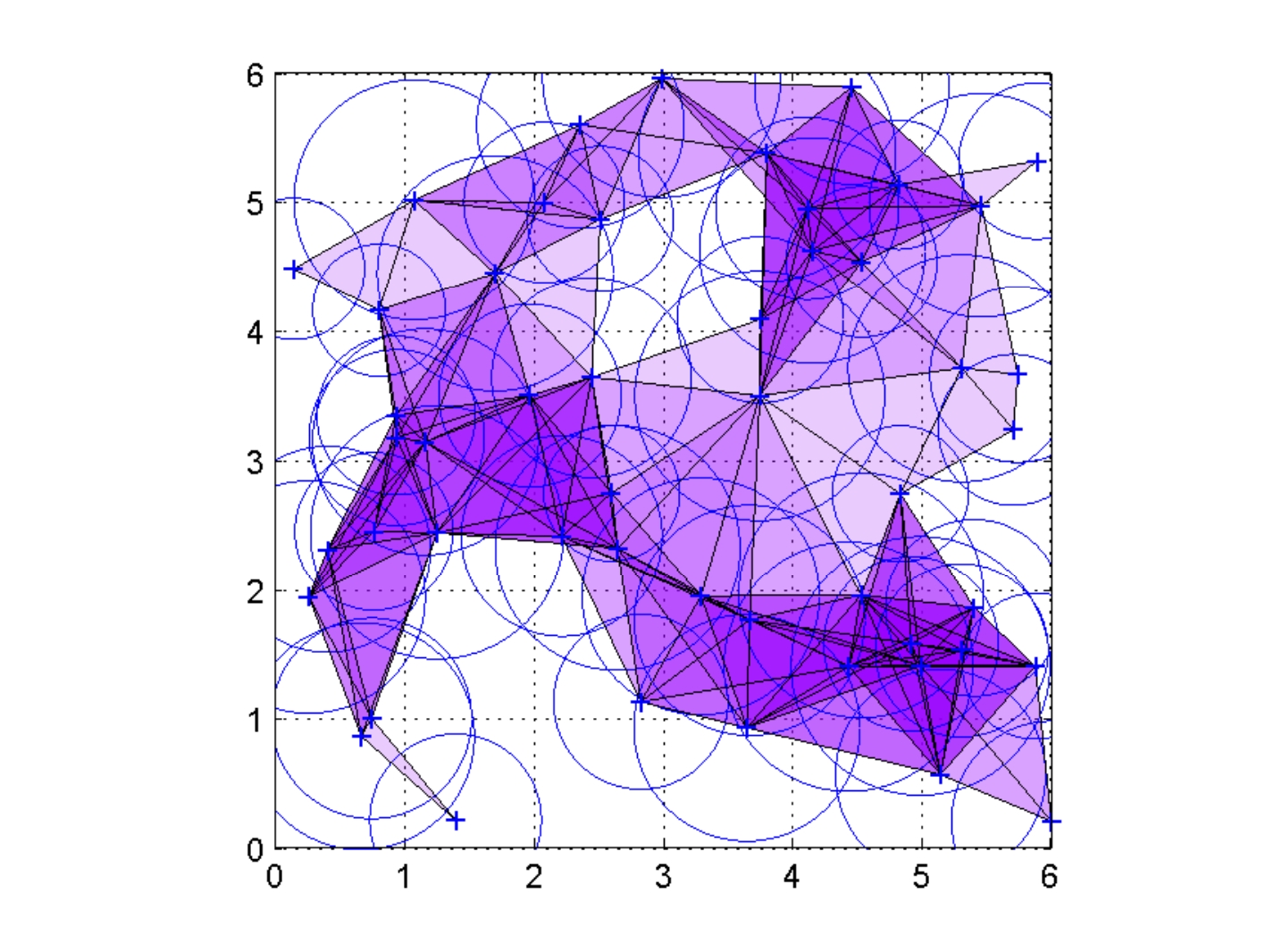}
\hfill
\caption{Random cells and their \v{C}ech complex.}
\label{fig:representation}
\end{figure}

In our simulations, the cells are deployed according to the Poisson point process on a square $6\times 6$. The density of cells varies from 1 (medium) to 2 (high). The radius of each cell can vary from 0.5 to 1. We use our algorithm to construct the generalized \v{C}ech complex for these cells up to dimension 2 and dimension 10. Note that, the generalized \v{C}ech complex built up to dimension 2 satisfies the requirement of almost applications in wireless networks. Our simulations are written in C++ language and executed on an Intel Core i7 2Ghz processor with 4GB of RAM.
The construction time of the \v{C}ech complex is listed in Table \ref{table:time}.
\begin{table}[H]%
\caption{Execution time (ms)}
\label{table:time}
\begin{center}
\begin{tabular}{|c|r|r|}
		\hline
		Density &$d_{\text{max}}=2$	&$d_{\text{max}}=10$\\
		\hline
		1	&2.31	&25.48\\
		1.5	&10.36	&580.42	\\
		2	&30.97	&10208.10 \\
		\hline
\end{tabular}
\end{center}
\end{table}
Figure \ref{fig:representation} shows the simulated cells with their representation by generalized \v{C}ech complex. 
In this Figure \ref{fig:representation}, the darker color indicates the higher dimension simplex, the lighter color indicates the lower dimension simplex. There is one coverage hole represented by the white space surrounded by colored simplices.

\section{Conclusion}
In this paper, we propose a centralized algorithm to build the generalized \v{C}ech complex. This complex is specified to analyze the coverage structure of wireless networks whose cells are different in size. This algorithm can build the minimal generalized \v{C}ech complex that is applied to wireless networks in polynomial time. Although this algorithm is designed for 2D space, it can be enhanced to be used in 3D space. Future work considers the design of the distributed release of this algorithm.



\bibliographystyle{IEEEtran}
\bibliography{IEEEabrv,myBib}
%




\end{document}